\documentclass{article}
\usepackage{graphicx} 
\usepackage{booktabs} 
\usepackage{multirow}
\usepackage{url}
\usepackage{tabularx}
\usepackage[margin=3.0cm]{geometry}

\title{Unmasking the Nuances of Loneliness: Using Digital Biomarkers to Understand Social and Emotional Loneliness in College Students}
\author{Malik Muhammad Qirtas}
\date{January 2024}

\begin{document}

\maketitle

\section{Abstract}
\textbf{Background:} Loneliness among students is increasing across the world, with potential consequences for mental health and academic success. To address this growing problem, accurate methods of detection are needed to identify loneliness and to differentiate social and emotional loneliness so that intervention can be personalized to individual need. Passive sensing technology provides a unique technique to capture behavioral patterns linked with distinct loneliness forms, allowing for more nuanced understanding and interventions for loneliness.

\textbf{Objectives:} This study aims to (1) utilize digital biomarkers extracted from passive sensing data to distinguish between socially and emotionally lonely students; (2) assess the predictive power of identified behavioral patterns in detecting and classifying levels of social and emotional loneliness; and (3) determine the most significant digital biomarkers used by ML models for predicting loneliness and it's two types.

\textbf{Methods:} To differentiate between social and emotional loneliness using digital biomarkers, our study included statistical tests, machine learning for predictive modeling, and SHAP values for feature importance analysis, revealing important factors in loneliness classification.

\textbf{Results:} Our analysis revealed significant behavioral differences between socially and emotionally lonely groups, particularly in terms of phone usage and location-based features , with machine learning models demonstrating substantial predictive power in classifying loneliness levels. The XGBoost model, in particular, showed high accuracy and was effective in identifying key digital biomarkers, including phone usage duration and location-based features, as significant predictors of loneliness categories.

\textbf{Conclusion:} This study underscores the potential of passive sensing data, combined with machine learning techniques, to provide insights into the behavioral manifestations of social and emotional loneliness among students. The identification of key digital biomarkers paves the way for targeted interventions aimed at mitigating loneliness in this population.

\section{Introduction}

In recent years, loneliness has emerged as a silent epidemic \cite{HHS2023}. Loneliness, as defined by Perlman, arises when an individual perceives a lack in their relationships, either in quality or quantity \cite{campbell2013loneliness}. While everyone may feel lonely at some point, it becomes a serious concern when it turns into a chronic condition \cite{martin2020differential}. Overall, loneliness is a deeply personal and emotional experience, characterized by dissatisfaction with one's social interactions and connections. It is important to note that loneliness is subjective; one person may live alone and feel content, while another may have a big social network and still feel lonely. Loneliness can negatively affect an individual's physical, emotional, and mental health \cite{hawkley2010loneliness,cacioppo2014social}. For example, research has shown that feeling lonely is linked to higher blood pressure \cite{hawkley2010loneliness}, a greater risk of developing cardiovascular diseases \cite{caspi2006socially}, reduced self-confidence \cite{cheng2002personality}, and a rise in symptoms of depression \cite{wei2005adult}.

While loneliness is often perceived as a singular emotional state, it is, in fact, a multifaceted experience \cite{weis1973experience}. Clinically, loneliness has been assessed using specific scales that delve into its multifaceted nature \cite{heinrich2006clinical,russell1982measurement}. One of widely used such scale is University of California Los Angeles (UCLA) scale which distinguishes loneliness into two primary factors: (1) emotional aspect, which highlights the absence of a close, intimate bond with another person, which can be deeply felt even when one is surrounded by others \cite{vaux1988social}, and (2) social aspect, which is perceived lack of social integration and belonging, often manifested when an individual has fewer interpersonal contacts or a smaller social network than desired \cite{russell1984social}. Studies have indicated that experiences of social and emotional loneliness do not always coincide. For instance, a person might have a robust peer network yet miss having a close friend, suggesting they feel emotional loneliness without necessarily feeling socially isolated \cite{maes2017intimate,qualter2002separateness}. This distinction is critical for a nuanced understanding of loneliness, as social and emotional loneliness may arise from different circumstances, have distinct physiological and psychological effects, and therefore may necessitate different intervention strategies \cite{lasgaard2011different}. Specifically, emotional loneliness has been linked with emotional challenges, including negative self-views, introverted behavior, and distorted views of relationships \cite{rokach2019psychological}. Recognizing the two forms not only enriches our comprehension of the overall phenomenon of loneliness but also empowers clinicians, educators, and policy-makers to craft more targeted and effective approaches to mitigate this pervasive issue.

College students represent a particularly salient group for the study of loneliness, given the unique set of challenges and transitions they encounter during this pivotal life stage. As they navigate the transition from adolescence to adulthood, students are often challenged by a new social environment, frequently far from their family support and their established social networks \cite{von2020development}. This challenge is coupled with the demands of academic stress, the pressure to conform to new social norms, and the challenge of establishing a sense of identity and belonging in an unfamiliar setting. Prior research has shed light on the prevalence of loneliness among this demographic; for instance, a study found that nearly 45\% of first-year college students reported feeling isolated and disconnected from their peers \cite{meehan2018really}. This is alarming, considering that such levels of loneliness have been correlated with a range of negative mental health outcomes, including increased rates of depression, anxiety, and stress. Therefore, understanding and addressing loneliness in the context of the college student experience is not only crucial for the well-being of the students themselves but also has broader implications for public health and educational systems.

In response to the need to better understand and address loneliness among college students, passive sensing data emerges as a promising avenue for research. Utilizing sensors embedded in ubiquitous devices such as smartphones and wearables like Fitbit, passive sensing data collection is non-intrusive and continuous, thereby offering a comprehensive and objective snapshot of an individual’s daily life \cite{torous2017new}. These data can include physical activity levels, sleep patterns, smartphone usage, and social interaction, as measured through call and message logs. Unlike self-reported surveys and questionnaires, which can be subject to recall bias and social desirability, passive sensing data provides an unfiltered lens into an individual’s behavior and well-being. In the context of loneliness, this technology has the potential to uncover subtle but telling signs of both social and emotional loneliness. Therefore, passive sensing data stands as a potentially revolutionary tool, not only for detecting loneliness in a timely and precise manner but also for shedding light on the nuanced ways in which loneliness manifests in daily life.

The objective of this study is to investigate the potential of passive sensing data to provide insight into the complex nature of loneliness that is prevalent among college students, highlighting the social and emotional aspects. Our goal is to determine the subtle differences between individuals who are socially and emotionally lonely, evaluate the predictive power of behavioral patterns in categorizing these loneliness types, and find the digital biomarkers that are most important for differentiating between these two types of loneliness. Our research is centered upon three fundamental questions.

\begin{enumerate}
  \item Can digital biomarkers extracted through passive sensing data differentiate between socially and emotionally lonely students?
  \item What is the predictive power of identified behavioral patterns to detect and classify levels of social and emotional loneliness?
  \item What digital biomarkers are most critical for predictive models in classifying loneliness and its types?
\end{enumerate}

\section{Methods}
\subsection{Dataset}
The dataset used in this study was gathered from a Carnegie-classified R-1 university in the United States \cite{xu2022globem}. For our research, we specifically utilize the dataset referred to as DS-2, which contains information from 218 undergraduate students. These students were initially recruited through email and social media outreach. This dataset covers a span of 10 weeks during the Spring quarter of 2019, ranging from late-March to mid-June, ensuring that potential seasonal effects were regulated. The sensing modalities used are location, phone usage, Bluetooth, WIFI, call logs, steps and sleep. The participants demographics are presented in Table-\ref{tab:demographicsTable}.

\begin{table}
\centering
\caption{Study Information and Participant Demographics. Gender acronyms: F: Female, M: Male, NB: Non-binary. Ethnic acronyms: A: Asian, B: Black or African American, H: Hispanic, N: American Indian/Alaska Native, PI: Pacific Islander, W: White, NA: Did not report. The \& symbol denotes participants who identified with multiple ethnicities.}
\label{tab:demographicsTable}
\begin{tabular}{l|l}
\toprule
\textbf{Category} & \textbf{Data} \\
\midrule
Participants & Total: 218 \\
 & Gender: F 111, M 107 \\
 & Ethnicity: A 102, B 6, H 10, N 2, PI 1, W 70, \\
 & A\&B 1, A\&W 16, H\&W 2, B\&W 2, \\
 & A\&H\&W 1, B\&H\&W 1, H\&N\&W 1, NA 3 \\
Ground Truth & Pre-study 10-items UCLA scale \\
 & Post-study 10-items UCLA scale \\
Sensor Data & Location, Phone Usage, Calls, Bluetooth Physical Activity, Sleep, Steps \\
\bottomrule
\end{tabular}
\end{table}

\subsection{Ethics and Privacy Considerations}
This study utilizes a pre-collected dataset that aims to enhance understanding of college students' daily behaviors and well-being through sensor data and self-reports. Ethical approval for this data collected was granted by the University of Washington's IRB (IRB number:STUDY00003244), and informed consent was obtained from all participants. To ensure data confidentiality, anonymization protocols were strictly adhered to, restricting direct identifiers to the primary data team. Moreover, data from participants who withdrew was promptly removed.

\subsection{Data Collection Tools}
The passive sensing data was collected using a smartphone application called AWARE \cite{ferreira2015aware}. The app is compatible with both iOS and Android based smartphones and continuously runs in the background without the user's active involvement. They also have used Fitbit to collect data for sleep and physical activities. For a more detailed understanding of physical activities and sleep patterns, the study combined readings from multiple sensors, including the accelerometer, gyroscope, microphone, and lightness sensor \cite{chen2013unobtrusive,wang2014studentlife}.

\subsection{Ground Truth for Loneliness Assessment}
For the evaluation of loneliness, the data collection study employed the revised 10-item UCLA loneliness scale \cite{knight1988some}. Participants were administered this questionnaire both before and after the study to assess the extent of their feelings of loneliness. Participants rated each of the 10 questions on a scale ranging from 1 (representing "never") to 4 (indicating "always"). Five of these items underwent reverse scoring, after which all the items were aggregated to generate a cumulative score. Given the four response options for each item, the overall score range for this 10-item questionnaire can range from 10 to 40, with higher scores indicating greater perceived loneliness.

For the purposes of our research, we have categorized the items of this scale into two distinct dimensions: social and emotional loneliness based on the criteria proposed in \cite{maes2022not,borges2008validacion}. Specifically, items that capture sentiments related to the lack of broader social interactions, such as the lack of companionship, feeling left out, and feeling isolated, are classified under social loneliness. This category comprises 5 items from the scale. In contrast, items that delve into feelings of intimate emotional disconnect, such as not feeling close to others or feeling that no one truly understands, are earmarked as indicators of emotional loneliness. This division encompasses the remaining 5 items from the scale. By distinguishing between these two dimensions, we aim to provide a nuanced approach to understanding loneliness in our dataset. The questions are given in the table \ref{tab:UCLADivision}

\begin{table}[ht]
    \centering
    \begin{tabularx}{\textwidth}{|c|X|}
        \hline
        \textbf{Type} & \textbf{Questions} \\
        \hline
        Emotional Loneliness & 
        1. How often do you feel that no one really knows you well? \\
        & 2. How often do you feel close to other people? (R) \\
        & 3. How often do you feel that there are people who really understand you? (R) \\
        & 4. How often do you feel that there are people you can turn to? (R) \\
        & 5. How often do you feel that people are around you but not with you? \\
        \hline
        Social Loneliness & 
        1. How often do you feel that you have a lot in common with the people around you? (R) \\
        & 2. How often do you feel that you feel left out? \\
        & 3. How often do you feel isolated from others? \\
        & 4. How often do you feel that there are people you can talk to? (R) \\
        & 5. How often do you feel that you lack companionship? \\
        \hline
    \end{tabularx}
    \caption{Division of 10-item UCLA scale into Emotional and Social Loneliness, 'R' indicates reverse scoring}
    \label{tab:UCLADivision}
\end{table}

The literature lacks a universally accepted threshold for determining loneliness scores, resulting in studies, including one by Cacioppo et al \cite{cacioppo2008loneliness}, proposing their own categories. In our study, participants were scored on two dimensions: social loneliness (referred to as social\_score) and emotional loneliness (referred to as emotional\_score), each ranging from 5 to 20. To establish these dimensions, we considered the potential answers to the questionnaire: 1 denoting "never," 2 as "rarely," 3 for "sometimes," and 4 for "often." A cumulative score of 10—achieved if a participant selected "rarely" for all 5 questions—suggested occasional feelings of social or emotional loneliness. Hence, we chose this as our cutoff score: scores at or below 10 reflected minimal to no feelings of loneliness, and scores above 10 implied moderate to high levels of loneliness. To categorize loneliness levels, we used the following approach:

\begin{itemize}
    \item Participants with a social\_score above 10 and an emotional\_score of 10 or less were labeled as "socially lonely."
    \item Those with a social\_score of 10 or less and an emotional\_score above 10 were labeled "emotionally lonely."
    \item Participants scoring above 10 on both scales were considered "both socially and emotionally lonely."
    \item Finally, those scoring 10 or below on both scales were categorized as "not lonely."
\end{itemize}

\subsection{Data Preprocessing}

In the preprocessing phase of our dataset, we applied data inclusion and exclusion criteria; specifically, we included only those students who had completed the post-study loneliness surveys. This results in the exclusion of participants who did not complete these surveys, underscoring our focus on obtaining a precise measure of perceived loneliness at the end of study. As a result, our final dataset comprised data from 205 students. The rationale behind choosing the post-study survey of loneliness, rather than pre-study survey, was driven by our aim to capture the students' actual feelings of loneliness as experienced throughout the study period. Academic stress, social adjustments, and other factors can significantly impact students' feelings of loneliness throughout the semester. By prioritizing the post-study measure, we aimed to capture the cumulative impact of these experiences on students' perceived loneliness. This provides a more holistic understanding of how the study period itself might have influenced their sense of connection and belonging.

To address the challenge of missing values and outliers within our dataset, a multi-step approach was adopted. Initially, we eliminated records with outliers by employing the Z-score method, which facilitated the identification and removal of anomalous data points based on their statistical deviation from the mean. Subsequent to outlier removal, missing continuous data were imputed using the median value for each respective feature, while the mode was used for imputing missing categorical data. This approach ensured a coherent and consistent dataset for subsequent analysis. Considering the varied nature of the algorithms employed in our study, we tailored our preprocessing steps to meet their specific requirements. For tree-based algorithms, which inherently accommodate varying scales of data, no feature scaling was applied. However, for other algorithms in our study, we implemented feature scaling using the 'standard scaler'. This normalization technique adjusted our numerical features to have a mean of zero and a standard deviation of one, thereby standardizing the data and optimizing it for analytical processing.

\subsubsection{Behavioural Features}
In our analysis of the passive sensing data, we used the RAPIDS framework to extract behavioral features. RAPIDS, a framework designed specifically for data preprocessing and digital biomarker extraction in passively collected datasets \cite{vega2020rapids}. This approach allowed us to extract a complete collection of behavioral features, including sleep and physical activity features, communication patterns, phone usage, and Bluetooth encounters. These features were critical for capturing the subtle parts of student behavior and daily routines that reflect their social and emotional states. Phone usage features such as total, maximum, and average duration of use revealed screen time patterns. Location features were retrieved to better understand spatial behavior, with features such as location variance, total distance traveled, and number of significant places visited. Movement features, such as average speed and radius of gyration, provided a picture of the mobility. Bluetooth interactions are tracked by the number of scans and unique devices encountered, revealing social exposure and interaction patterns in the person's vicinity. Call logs provided insight into communication patterns, including indicators such as the frequency of missed calls and unique contacts. Physical activity was tracked using step count analytics, which provided an indication of daily activity levels. Finally, sleep patterns were inferred using a variety of indicators, including total time sleeping and awake, as well as sleep efficiency, providing a full picture of rest and circadian rhythms. Collectively, these factors created a rich, multidimensional dataset that allowed for an in-depth investigation of behaviors related with social and emotional loneliness in students. For a complete list and explanation of all the features extracted using the RAPIDS framework, please see the RAPIDS documentation \cite{vega2020rapids}. 

\subsection{Analyzing Loneliness Profiles: Basic Statistics}
To investigate the different intricate aspects of loneliness among the participants in our study, we utilized a comprehensive statistical methodology to gain insight into the broad and subtle aspects of loneliness as measured by the UCLA loneliness scores. We initiated our analysis by computing fundamental descriptive statistics for the overall UCLA loneliness scores, specifically calculating the mean, median, Q1, Q3, and standard deviation (SD). Following our descriptive analysis, we proceeded with a more comprehensive classification of participants according to their total loneliness scores, dividing them into primary categories of low and high loneliness. Then we computed the same set of descriptive statistics specifically for different loneliness profiles, namely overall loneliness, social loneliness, and emotional loneliness. 

We also conducted a distribution analysis to categorize the prevalence and types of loneliness among participants. Participants were classified into low and high loneliness categories based on their overall UCLA scores. We categorized the overall UCLA loneliness score into low (<=20) and high categories (>20). We calculated the number of participants in each category, enabling us to quantify the prevalence of significant loneliness within our sample. We further analyzed the distribution of participants across four distinct categories - "socially lonely", "emotionally lonely", "both socially and emotionally lonely", and "not lonely".

\subsection{Differentiating Social and Emotional Loneliness}
To differentiate between social and emotional loneliness using digital biomarkers, our study conducted a statistical analysis. Initially, we assessed the normality of the distribution of digital biomarkers data for both socially and emotionally lonely student groups. This was conducted using the Shapiro-Wilk test \cite{royston1983some}, a robust method for testing normality, particularly effective for small sample sizes like ours (24 participants in the socially lonely group and 19 in the emotionally lonely group). The choice of this test was due to its sensitivity in detecting deviations from a normal distribution in small datasets.

Given the potential non-normal distribution of the data, as indicated by the Shapiro-Wilk test, we employed a two-sided Mann–Whitney U test to compare the digital biomarker distributions between the socially and emotionally lonely groups. This non-parametric test was selected for its ability to compare differences between two independent samples without the assumption of normal distribution, making it ideal for our analysis. The test aimed to determine if there were statistically significant differences in the digital biomarker profiles between socially lonely and emotionally lonely groups.

To quantify the magnitude of the differences observed between the two groups using the Shapiro-Wilk test, we applied the bootstrapping method for effect size calculation. A total of 10K bootstrap samples were used to estimate the distribution of the effect size and resampled biomarkers with replacement, which provides a more robust measure in the context of our small sample sizes. We selected Cohen's d as our effect size metric, given its suitability for comparing means between two groups. Alongside the point estimate, a 95\% confidence interval was computed from the bootstrap distribution to provide an estimate of the precision of the calculated effect size.

\subsection{Predictive Modeling for Loneliness Classification}
To evaluate the predictive power of identified digital biomarkers for detecting and classifying levels of social and emotional loneliness, our methodological approach incorporates advanced machine learning algorithms. The dataset comprised instances labeled as 'socially lonely,' 'emotionally lonely,' 'both lonely,' or 'not lonely.' These labels were derived from validated measures and were used as target variables for the classification models. Given the varied prevalence of loneliness categories, we identified class imbalances that could potentially bias our predictive models. To address this, we employed the Synthetic Minority Over-sampling Technique (SMOTE) \cite{chawla2002smote}. This technique generated synthetic examples in the feature space for underrepresented classes, thereby balancing the class distribution and improving the generalizability of classification models.

For the classification task, we selected a number of machine learning algorithms known for their efficacy in multi-class classification scenarios: Support Vector Machine (SVM), XGBoost, Random Forest, and K-Nearest Neighbors (KNN). For each model, hyperparameters were tuned using grid search or random search techniques to optimize performance. To evaluate the models, we applied nested cross-validation, specifically leave-one-person-out cross-validation (LOOCV). This rigorous form of validation involves using data from all participants minus one for training, and then testing the model on the remaining participant. This process iterates until each participant has been used as the test set once. LOOCV is particularly suited for datasets containing personal data, ensuring that the evaluation of the model's performance reflects its ability to generalize across unseen data.

The primary metrics for assessing the performance of each model were accuracy, precision, recall, and F1 score. These metrics were calculated for each loneliness category, providing a detailed view of the model's predictive power. Accuracy offered a global view of model performance, while precision, recall, and the F1 score provided insights into the model's ability to predict each specific class. This multi-faceted evaluation framework allowed us to comprehensively understand the strengths and limitations of our predictive models.

\subsection{Features Importance Analysis}
To determine the most important features in distinguishing between the different loneliness types, we employed SHapley Additive exPlanations (SHAP) values \cite{lundberg2020local}. SHAP values offer a powerful framework for interpreting predictions of machine learning models, providing insights into the contribution of each feature to the model's output. Once the classification models were trained, we used the SHAP library to compute the SHAP values for each feature across all data points in our validation set. This process outputs a SHAP value for each feature for each prediction, reflecting the feature's impact on the model's output. To synthesize these results, we averaged the absolute SHAP values across the validation set for each feature, thus obtaining a global measure of feature importance. We then ranked the features based on their average SHAP value, giving us a clear picture of which features were most significant in predicting loneliness classes.

\section{Results}

\subsection{Insights into Loneliness Dimensions among Participants}
The overall UCLA loneliness scores were a focal point in exploring loneliness among our participants. We observed an average score slightly above the established loneliness threshold, with a mean score of 21.64, a median of 21, and an interquartile range spanning from 18 (Q1) to 25 (Q3). The variability in loneliness scores, as denoted by the standard deviation, was 5.32. When categorizing participants based on these scores, we found that a little over half, 55.12\% (113 out of 205), were categorized under high loneliness, whereas 44.88\% (92 out of 205) fell into the low loneliness category.

Diving deeper into the dimensions of loneliness, the social loneliness scores painted a picture of the participants' interpersonal disconnect. The mean score stood at 10.93, with a median score of 11 and an interquartile range of 9 (Q1) to 13 (Q3). The distribution of these scores had a standard deviation of 2.736. Emotional loneliness revealed a mean score of 10.71, a median score of 11, and an interquartile range of 9 (Q1) to 13 (Q3). The standard deviation for these scores was slightly higher at 2.905.

When categorizing our cohort based on their social and emotional loneliness scores, our analysis yielded intriguing insights. A smaller subset, 11.71\% (24 out of 205) were deemed socially lonely, while 9.27\% (19 out of 205) were identified as emotionally lonely. Notably, a significant 42.44\% (87 out of 205) participants felt socially and emotionally isolated, revealing an intertwined nature of these loneliness dimensions. Meanwhile, 36.59\% (75 out of 205) felt neither socially nor emotionally isolated, a silver lining in our findings.

\subsection{Statistical Differences for Social and Emotional Loneliness}
\begin{figure}[htbp]
\centering
\includegraphics[scale=0.84, page=1]{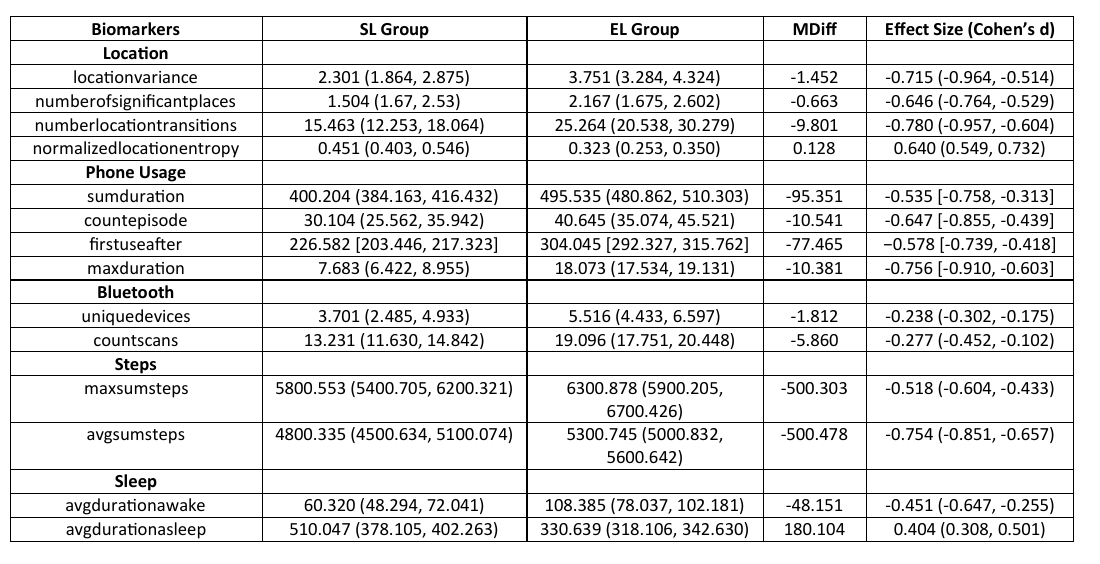}
\caption{Mean Difference and Effect Sizes of biomarkers between socially and emotionally lonely groups}
\label{fig:pdf_rq1}
\end{figure}

The Shapiro-Wilk test for normality check indicated a non-normal distribution of data, leading us to employ two-sided Mann-Whitney U test non-parametric test for calculating differences between loneliness groups. Table~\ref{fig:pdf_rq1} presents the statistically significant (p < 0.05) mean differences and their effect sizes for socially and emotionally lonely groups. 

Results indicate that the Socially Lonely (SL) group experienced less variance in their locations [Mean: 2.301 (CI: 1.864, 2.875)] compared to the Emotionally Lonely (EL) group [Mean: 3.751 (CI: 3.284, 4.324)], with a mean difference (MDiff) of -1.452 and an effect size (Cohen’s d) of -0.715 (CI: -0.964, -0.514). Additionally, the SL group visited fewer significant places and had fewer location transitions than the emotionally lonely group. In terms of phone usage, the SL group [Mean: 400.204 (CI: 384.163, 416.432)] used their phones (overall generic phone usage) for a shorter duration compared to the EL group [Mean: 495.535 (CI: 480.862, 510.303)], with a significant MDiff of -95.351 and a small effect size of -0.535 (CI: -0.758, -0.313). The frequency of phone usage episodes and the duration before the first use after waking up were also lower in the SL group. The use of Bluetooth devices revealed a lower engagement for the SL group [Mean: 3.701 (CI: 2.485, 4.933)] relative to the EL group [Mean: 5.516 (CI: 4.433, 6.597)], with a MDiff of -1.812 and a moderate effect size of -0.238 (CI: -0.302, -0.175). Similarly, the SL group performed fewer Bluetooth scans as compared to the emotionally lonely group. The SL group reported fewer maximum and average steps taken [Max Mean: 5800.553 (CI: 5400.705, 6200.321); Avg Mean: 4800.335 (CI: 4500.634, 5100.074)] compared to the EL group [Max Mean: 6300.878 (CI: 5900.205, 6700.426); Avg Mean: 5300.745 (CI: 5000.832, 5600.642)], with effect sizes of -0.518 and -0.754 respectively. Significant differences were also observed in sleep patterns, with the SL group averaging less time awake [Mean: 60.320 (CI: 48.294, 72.041) minutes] and more time asleep [Mean: 510.047 (CI: 378.105, 402.263) minutes] compared to the EL group [Awake Mean: 108.385 (CI: 78.037, 102.181) minutes; Asleep Mean: 330.639 (CI: 318.106, 342.630) minutes]. The effect sizes were -0.451 for awake duration and 0.404 for asleep duration.

\subsection{Predictive Accuracy of Digital Biomarkers in Loneliness Categories}
\begin{figure}[htbp]
\centering
\includegraphics[scale=0.84, page=1]{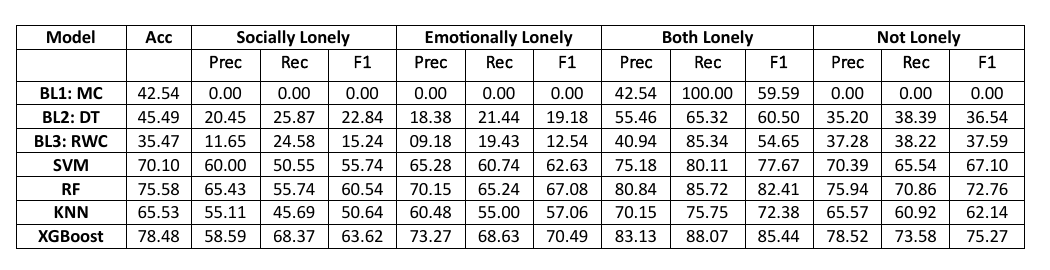}
\caption{Performance of classification models for loneliness prediction. The model performances are compared with three baseline
classifiers (BL1:MC = BaseLine 1 Majority Class; BL2:DT = BaseLine 2 Decision Tree, and BL3:RWC = BaseLine 3 Random Weighted Classifier; SVM = Support Vector Machine, RF = Random Forest, KNN = K-Nearest Neighbors). All values are reported as percentages.}
\label{fig:ml_classification}
\end{figure}

The predictive performance for 4 loneliness categories using machine learning classifiers, along with the 3 baseline models, are presented in Table~\ref{fig:ml_classification}. The baseline models established a foundational understanding of the dataset's inherent patterns. The Majority Class (BL1: MC) classifier, which categorically predicted the 'Both Lonely' category, achieved an accuracy of 42.54\%, reflecting the prevalence of this class in the dataset. Despite its simplicity, the Decision Tree (BL2: DT) demonstrated a modest improvement in accuracy (45.49\%). The Random Weighted Classifier (BL3: RWC) yielded an accuracy of 35.47\%, the lowest among the models, suggesting that a strategy based solely on class distribution is insufficient for this task.

The XGBoost model was the standout performer, with the highest overall accuracy of 78.48\%. Notably, it achieved the highest recall in the 'Both Lonely' category (88.07\%) and the highest F1 scores in the 'Both Lonely' (85.44\%) and 'Not Lonely' (75.27\%) categories, suggesting a strong capability to identify participants with both social and emotional loneliness characteristics, as well as those without loneliness. The Support Vector Machine (SVM) reached an accuracy of 70.10\%, with particularly strong performance in predicting the 'Both Lonely' and 'Not Lonely' categories. The Random Forest (RF) model demonstrated the highest precision and F1 scores across the 'Both Lonely' and 'Not Lonely' categories and achieved the second-highest overall accuracy at 75.58\%. K-Nearest Neighbors (KNN) showed balanced performance, with an overall accuracy of 65.53\%.

\subsection{Important Features for Loneliness Classification}

The SHAP value analysis for the XGBoost model indicates the relative importance of different features in predicting the loneliness categories of 'socially lonely,' 'emotionally lonely,' 'both lonely,' and 'not lonely.' Features related to phone usage, specifically maximum duration of use, and location-based metrics such as the maximum length of stay at clusters and location variance show a substantial impact across all classes, with particularly high influence in the 'both lonely' category. Notably, the number of significant places and the number of location transitions are crucial in differentiating between the loneliness classes, suggesting that mobility patterns could be a key indicator of social and emotional loneliness. Similarly, standard deviations in phone usage duration and steps also show a significant mean SHAP value, highlighting variability in daily routines as an important factor. The SHAP values decrease as we move down the feature list, but even those with lower values, like the average duration in bed, still contribute to the model's ability to discern between loneliness categories.

In our Random Forest model, the SHAP value analysis revealed a distinct hierarchy of feature importance across the four loneliness categories. Location-based features such as variance and entropy emerged as highly influential, with the greatest impact on the 'both lonely' category, signifying that geographic movement patterns are strong indicators of loneliness levels. Phone usage patterns, such as the number of unlock episodes (countepisode) which reflects the frequency of phone engagement, and the longest duration of any unlock episode (maxduration), which signifies the extent of the most extended interaction with the phone, were also significant. The maximum duration of phone use, indicating prolonged use in a single session, was significantly different in the 'socially lonely' group, suggesting a distinct pattern that could be predictive of social loneliness. Notably, location feature  had a considerable effect on the 'emotionally lonely' category, highlighting the potential link between the amount of time spent in habitual locations and emotional loneliness. The model indicated that overall phone use time and total distance traveled play a moderate role in classification, suggesting that overall phone use and movement within an environment are informative for loneliness prediction. Interestingly, sleep-related features like average duration of sleep showed significant importance in the 'not lonely' category.

\begin{figure}
    \centering
    \begin{minipage}{.72\textwidth}
        \includegraphics[width=\linewidth]{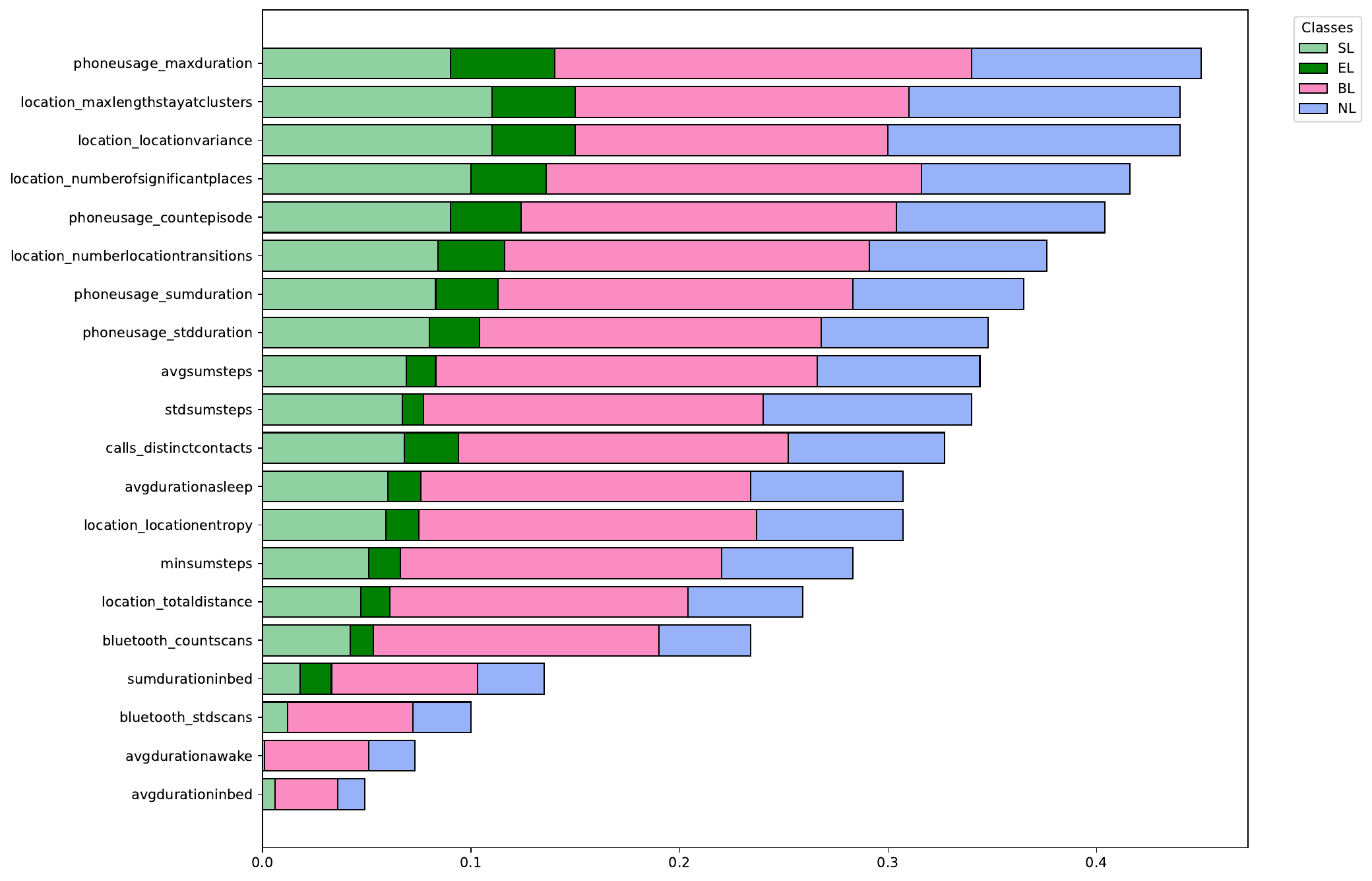}
        \caption{Mean Feature Importance using SHAP for XGBoost model}
        \label{fig:shap_XGBoost}
    \end{minipage}%
    \hfill
    \begin{minipage}{.72\textwidth}
        \includegraphics[width=\linewidth]{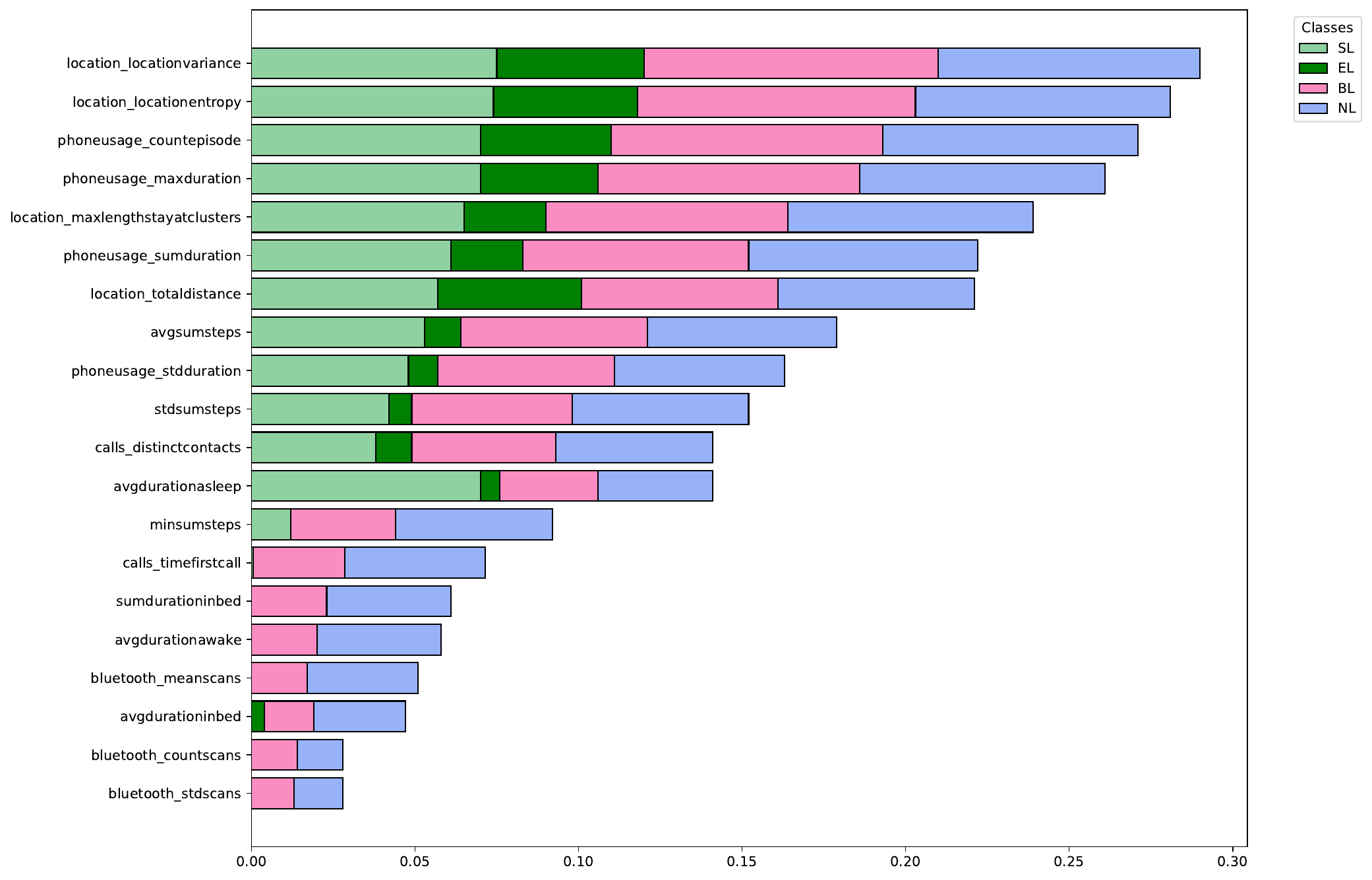}
        \caption{Mean Feature Importance using SHAP for Random Forest model}
        \label{fig:shap_randomforest}
    \end{minipage}
\end{figure}

\section{Discussion}
The main aims of this study were the investigation of digital biomarkers, their prediction capabilities, and their importance in differentiating across loneliness types. The analysis of passive sensing data to differentiate between socially and emotionally lonely students revealed significant statistical findings, underscoring the potential of digital biomarkers in mental health research. The observed differences in digital biomarkers between the Socially Lonely (SL) and Emotionally Lonely (EL) groups provide evidence that passive sensing can capture distinct patterns associated with these two forms of loneliness. The greater variance in location data and higher engagement with significant places and transitions in the EL group implies a pattern of seeking social connection outside of their usual environments, probably as a compensating method for emotional loneliness. In contrast, the SL group's lower mobility and fewer significant places visited may indicate a lack of interest or opportunity to participate in social interactions, which is a sign of social loneliness. This difference in the physical world relationship between the two groups shows the subtle ways in which loneliness manifests. While these observations highlight potential behavioral differences between the groups, it's important to note that our current data cannot definitively establish a causal relationship. Further research, including qualitative studies to understand the motivations and experiences behind these behavioral patterns, is necessary to explore the complex interplay between loneliness and real-world behavior. The findings related to phone usage patterns further differentiate the two types of loneliness. The EL group's higher phone use, both in terms of time and frequency, might indicate a preference for digital communication over direct social engagement, potentially due to feelings of emotional loneliness. The SL group's reduced engagement with their phones might be attributed to a different social context or a decreased desire for compensating digital contact. The Bluetooth data, indicating fewer unique devices encountered and less frequent scans by the SL group, suggests a lower level of social activity or physical proximity to others, aligning with the characteristics of social loneliness. In contrast, the EL group's higher engagement may reflect their participation in more social situations but without emotional connection. Significant differences in physical activity and sleep patterns between the groups provide additional insights into the physiological and behavioral impacts of loneliness. The reduced physical activity in the SL group might contribute to or result from social withdrawal, while the sleep pattern differences, with the SL group reporting more sleep, might indicate depressive symptoms or a lack of daytime engagement, needing less sleep. These findings have important implications for understanding the complex nature of loneliness and its impact on student behavior and well-being. They suggest that passive sensing technology can differentiate between social and emotional loneliness and offer insights into the behavioral patterns associated with each. These findings can be utilised for targeted interventions, such as promoting social engagement for socially lonely students or addressing emotional needs and connectivity for emotionally lonely students. To maximize the effectiveness of such interventions, it is crucial to first gain a deeper understanding of the motivations and experiences behind the behavioral differences we observed. Qualitative research could illuminate why individuals engage in specific patterns and how they perceive the connection between their behaviors and feelings of loneliness.

The results from the machine learning analysis of behavioral patterns to detect and classify loneliness types reveal significant insights into the predictive power of digital biomarkers. The standout performance of the XGBoost model, with an overall accuracy of 78.48\%, underscores the potential of advanced ensemble methods in handling complex, multi-dimensional datasets like those derived from passive sensing. The high recall and F1 scores for the 'Both Lonely' category indicate XGBoost's robustness in identifying nuanced states of loneliness, a critical capability for interventions aimed at mitigating both social and emotional loneliness. This analysis emphasizes the practical implications for developing targeted interventions. Interventions can be more finely tailored by accurately classifying individuals into specific loneliness categories, potentially enhancing their effectiveness. For instance, individuals identified as 'Both Lonely' might benefit from interventions addressing both social aspects and emotional support, while those classified as 'Not Lonely' might be targeted for preventive measures.

The feature importance analysis analysis from the ML models provides critical insights into the digital biomarkers most effective in distinguishing between socially lonely and emotionally lonely participants. This differentiation is crucial for understanding the nuanced ways in which loneliness manifests and informs targeted interventions. Notably, digital biomarkers associated with phone usage patterns emerged as strong predictors across loneliness categories. This shows that the level of phone use may be compensatory behavior for a lack of social engagement or an escape strategy for individuals who are lonely. The strong influence of location-based features, such as maximum duration of stay at clusters and location variation, emphasizes the importance of physical mobility and routine in the feeling of loneliness. High variance in location and fewer significant places visited may indicate a lack of social engagement or a stable social routine, factors often associated with social loneliness. Moreover, the analysis highlights the importance of variability in daily routines, as captured by standard deviations in phone usage duration and steps, suggesting that irregular patterns could signal disrupted social or emotional well-being. The importance of sleep-related features, particularly the average sleep duration, points to the potential role of rest in overall mental health and its connection to loneliness. This finding aligns with existing research on the relationship between sleep and loneliness, further supporting the idea that sleep disruptions may be a contributing factor or consequence of loneliness \cite{hom2020meta,griffin2020loneliness,doane2014associations}. Interestingly, the predictive power of these biomarkers varies between the two loneliness categories, with some features, like location transitions and significant places, being more indicative of social loneliness, while others, such as the maximum length of stay at clusters, more closely align with emotional loneliness. This differentiation is important for developing targeted interventions, where understanding the specific type of loneliness experienced can guide more personalized support strategies.

Our study, although novel in its approach of distinguishing between social and emotional loneliness using digital biomarkers, has some limitations that call for more research. First, our results' generalizability is limited by the size and diversity of our sample. The participants, who are college students, form a somewhat homogeneous group with possibly comparable daily routines and psychological challenges. This homogeneity limits the scope of our research and its relevance to a broader population. Future research might benefit from a larger and more diversified population to confirm and generalize these results. Furthermore, the dataset used in our study, although rich in digital biomarkers, may not capture all of the characteristics associated with loneliness. Personal relationships, age, family history, major life events, mental health history, and individual coping techniques were not taken into consideration. These factors have an important part in determining an individual's loneliness experience and may have a major impact on the results of such study. Additionally, it's important to consider potential changes to the validity of the loneliness scale when differentiating between social and emotional loneliness. As a result, a more holistic approach, encompassing both emotional and psychological elements, is needed for a thorough understanding of loneliness.

While our study contributes valuable preliminary insights into the differentiation of social and emotional loneliness through digital biomarkers, it also highlights the need for broader, more inclusive research. Future work should aim to overcome these limitations, expanding the scope and depth of loneliness research to foster more effective detection, understanding, and intervention strategies. Further research could look into the use of qualitative data, such as self-reported loneliness assessments, personal interviews, or diaries, to add to the quantitative findings from passive sensing. This mixed-methods approach has the potential to provide deeper insights into the subjective experiences of loneliness, resulting in a more nuanced understanding of different types of loneliness. Furthermore, the advancement of machine learning models and analytical approaches provides an opportunity to improve the prediction and categorization of loneliness. The insights gained from this study have the potential to transform our approach to addressing the growing problem of loneliness. By leveraging passive sensing technology, we can not only identify those at risk but also differentiate between social and emotional loneliness, enabling the development of highly targeted and personalized interventions. This focus on early detection through passive sensing paves the way for timely interventions aimed at fostering social connection, supporting emotional well-being, and promoting overall mental health for those struggling with loneliness.

\end{document}